# IMPLEMENTATION OF DYMO ROUTING PROTOCOL

Anuj K. Gupta[1], Harsh Sadawarti[2] and Anil K. Verma[3]

[1]Associate Prof. & Head, CSE Dept., RIMT IET, Mandi Gobindgarh, PB, India
anujgupta@rimt.ac.in
[2]Director & Professor , CSE Dept., RIMT IET, Mandi Gobindgarh, PB, India
harshsada@yahoo.com
[3]Associate Professor, CSE Dept., Thapar University, Patiala, PB, India
akverma@thapar.edu

**_ABSTRACT_**

*Mobile ad hoc networks communicate without any fixed infrastructure or ant centralized domain. All the nodes are free to move randomly within the network and share information dynamically. To achieve an efficient routing various protocols have been developed so far which vary in their nature and have their own salient properties. In this paper, we have discussed one of the latest protocols i.e. Dynamic Manet on demand (DYMO) routing Protocol, implemented and analysed its performance with other similar protocols against different parameters. Finally a comparison has been presented between all of them.*

## 1. INTRODUCTION

A Mobile Ad hoc NETwork (MANET) is a collection of two or more autonomous nodes which communicate with each other without any centralized administering node. MANETs possess few salient features such as dynamic topology, limited storage and bandwidth which make them attractive for certain applications but at the same time pose challenges to route packets efficiently and accurately to a particular destination. These types of networks lack in fixed topology hence they are often known as infrastructureless networks, where each node has the capacity to work both as router or host or both. Routing in MANET is a challenging task and has gained a remarkable attention from researchers worldwide. It has been observed from the literature survey that none of the existing protocols is the best that justifies the characteristics and is suitable to perform an efficient routing. Researchers strive to uncover the efficiency of existing routing protocol by enhancing its performance in terms of various metrics like throughput, end to end delay, packet delivery ratio, etc.

## 2. ROUTING PROTOCOLS

A routing protocol is a mechanism by which the network traffic is directed and transported through the network from source to destination. In addition to this, these routing protocols may need to provide different levels of Quality of Service (QoS) to support different types of applications and users. In traditional wired networks two main conventional algorithmic strategies were mainly used: link-state and distance vector algorithms. In link-state routing, each node maintains up-to-date information of the network by periodically broadcasting the link-state costs of its neighbouring nodes to all other nodes using flooding technique. When each node receives an update packet, they update their link-state information by applying a shortest-path algorithm to choose the next hop node for each destination [4].

To overcome the problems associated with the link-state and distance-vector algorithms a number of routing protocols have been proposed. These Routing Protocols are broadly classified into two categories: flat routing and hierarchical routing. There are further two classes in flat routing: reactive (on-demand) and proactive (table-driven) [5]. The protocols that are a combination of both reactive and proactive characteristics are referred to as hybrid, which are based on hierarchical routing. The table-driven or global ad hoc routing approach is similar to the connectionless approach of forwarding packets, with no regard to when and how frequently routes are desired. But this is not in the case of on-demand routing protocols [6].

## 3. Dynamic MANET On demand Routing Protocol (DYMO)

DYMO routing protocol has been proposed by Perkins & Chakeres [3] as advancement to the existing AODV protocol. It is also defined to as successor of AODV or ADOVv2 and keeps on updating till date. DYMO operates similar to its predecessor i.e. AODV and does not add any extra modifications to the existing functionality but operation is moreover quite simpler. DYMO is a purely reactive protocol in which routes are computed on demand i.e. as and when required. Unlike AODV, DYMO does not support unnecessary HELLO messages and operation is purely based on sequence numbers assigned to all the packets. It is a reactive routing protocol that computes unicast routes on demand or when required. It employs sequence numbers to ensure loop freedom. It enables on demand, multi-hop unicast routing among the nodes in a mobile ad hoc network. The basic operations are route discovery and maintenance. Route discovery is performed at source node to a destination for which it does not have a valid path. And route maintenance is performed to avoid the existing obliterated routes from the routing table and also to reduce the packet dropping in case of any route break or node failure.

### 3.1 Route Messages

DYMO implements three messages during the routing operation namely Route Request (RREQ), Route Reply (RREP) and Route Error (RERR).

1. RREQ message is used by source node to discover a valid route to a particular destination node.
2. RREP message is used to set up a route between destination node and source node, and all the intermediate nodes between them.
3. RERR message is used to indicate a invalid route from any intermediate node to the destination node.

Besides this the DYMO protocol mandates each node to maintain an unsigned unique integer called as “sequence number” which guarantees the orderly delivery of packets to the destination and maintain loop-free routes similar to that in AODV and DSDV. Sequence numbers allow the nodes to evaluate the freshness of routing information.

### 3.2 Route Discovery

The DYMO route discovery is very similar to that of AODV except for the path accumulation feature. Figure 1 shows the DYMO route discovery process. If a source has no route entry to a destination, it broadcasts a RREQ message to its immediate neighbours. If a neighbour has an entry to the destination, it replies with an RREP message else it broadcasts the RREQ message.

While broadcasting the RREQ message, the intermediate node will attach its address to the message. Every intermediate node that disseminates the RREQ message makes a note of the backward path. With respect to figure (()), source node 1 wants to communicate with destination node 10. It generates a RREQ packet which contains its own address, sequence number, hop count, destination address, and broadcasts it on the network. Each intermediate node having a valid path to the destination keeps on adding its address and sequence number to the RREQ packet as shown with nodes 2 and 6, till destination is reached. The source node waits for a RREP message. The Destination replies with RREQ message. A similar path accumulation process takes place along the backward path. This makes sure that the forward path is built and every intermediate node knows a route to every other node along the path. If source does not receive RREP within a specified TTL value, RREQ may be resend.

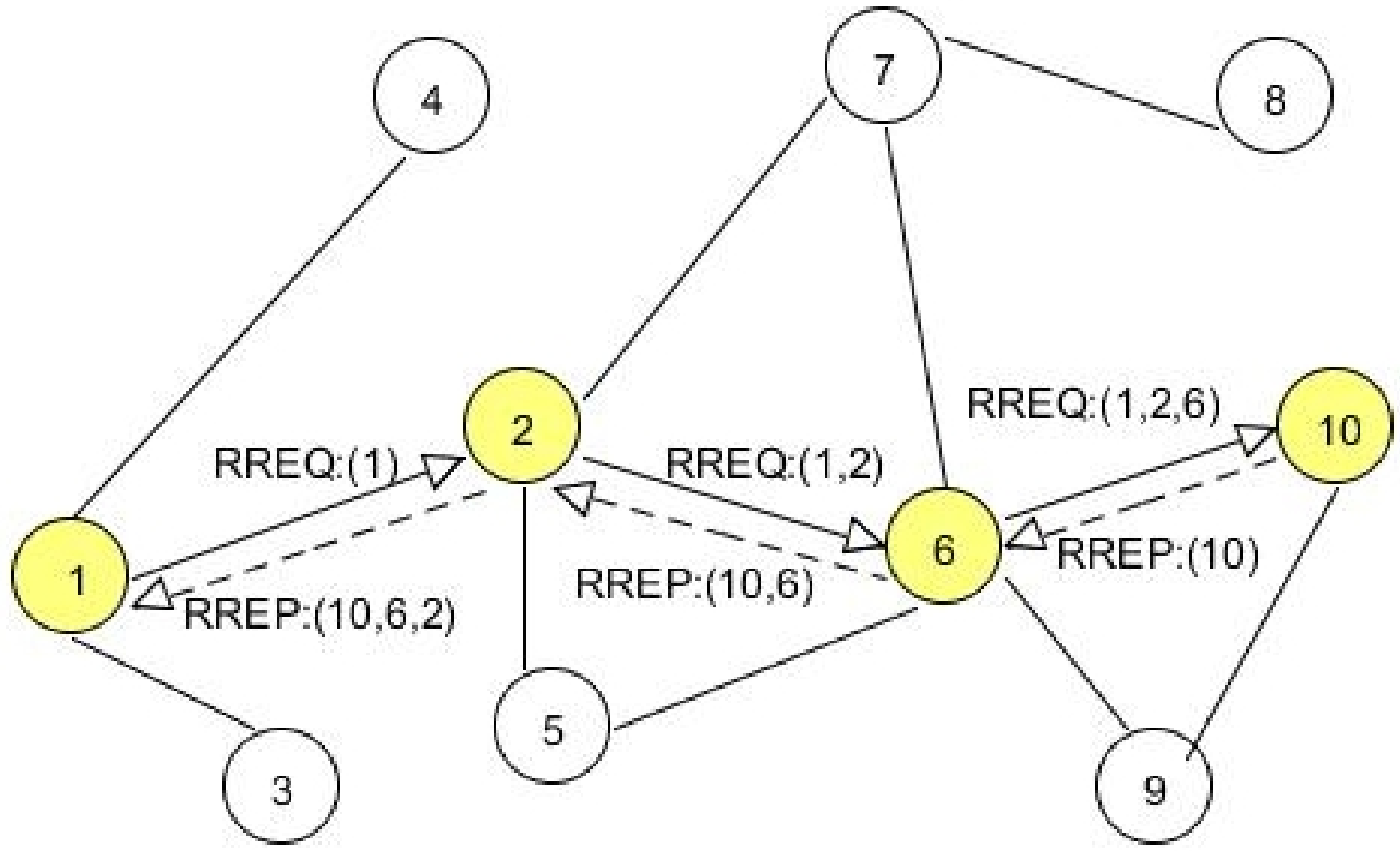

Figure 1. DYMO Route Discovery

Each node maintains a unique sequence number in order to avoid loops in the route and also to discard the stale packets if any. Every time a RREQ is sent, the router updates its sequence number. If the incoming packet has a same or inferior sequence number, the information is discarded. Messages with superior sequence numbers are updated in the routing table. If the sequence number associated with the incoming route is the same as the node sequence number then a loop is possible. In such case, the incoming packet is discarded. One of the special features of DYMO is that it is energy efficient. If a node is low on energy, it has the option to not participate in the route discovery process. In such a case, the node will not forward any of the incoming RREQ messages. It however will analyze the incoming RREP messages and update its routing tables for future use

### 3.3 Route Maintenance

During the routing operations each node has to continuously monitor the status of links and maintain the latest updates within the routing tables. The route maintenance process is actually accomplished with the help of RERR messages. The RERR message must be generated by a node if and when a link to any other node breaks. The generating node multicasts the RERR message to only those nodes which are concerned with the link failure. Upon reception of a RERR message, the routing table is updated and the entry with the broken link is deleted. If any of the nodes face a packet to the same destination after deletion of the route entry, route discovery process needs to be initiated again.

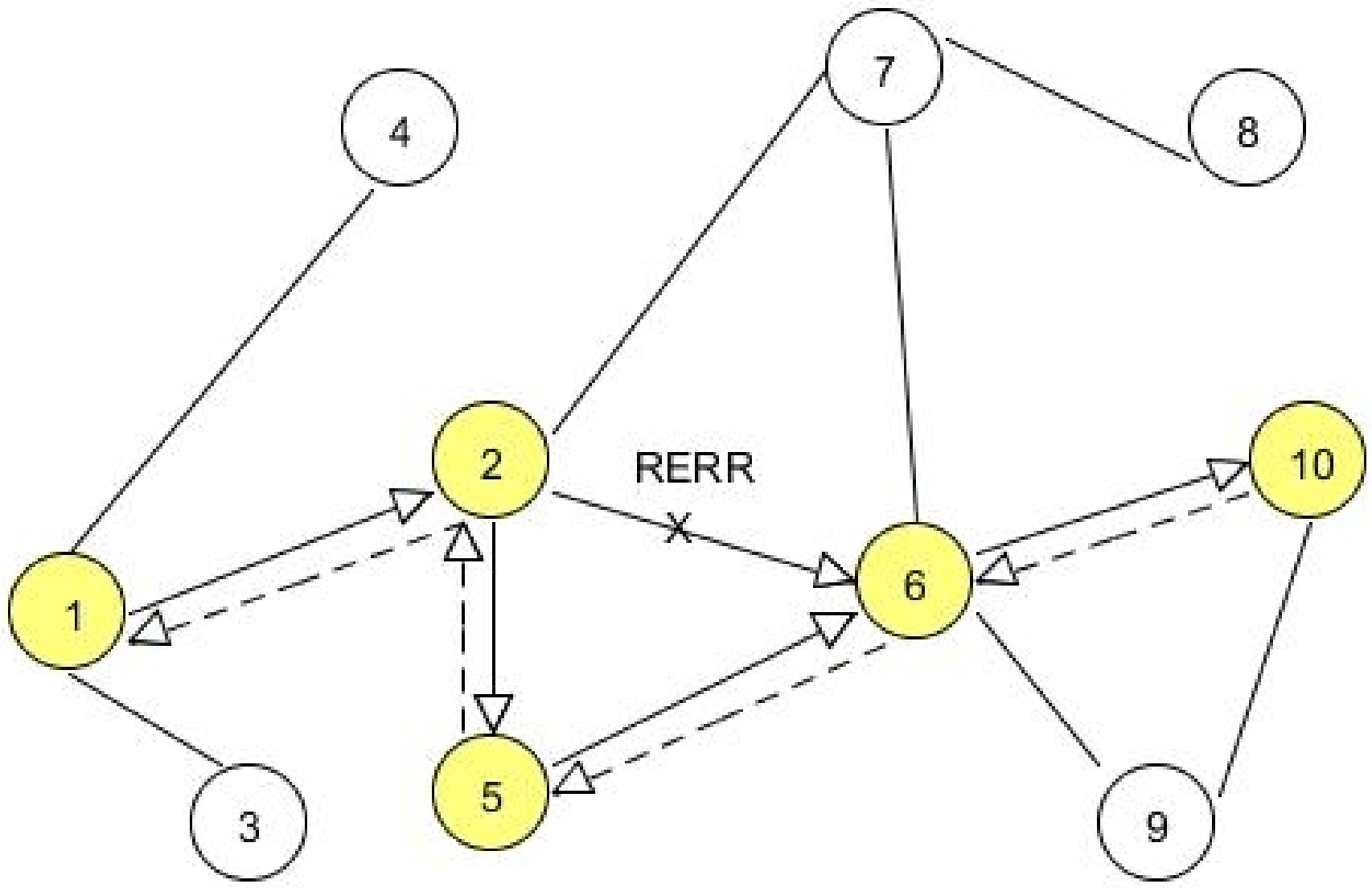

Figure 2. DYMO Route Maintenance

As seen from the figure 2, node 2 has received a packet that needs to go to node 6, but the route from node 2 to node 6 is found broken. In this case a RERR message is generated by node 2 and forward towards the source node 1. All the intermediate nodes on the path instantly update their routing table entries with the new updated information regarding link failure and new route changes. Now the packets will be forwarded from node 2 towards node 5 and then to node 6 and lastly to node 10 so as to reach the destination.

### 3.4 Advantages and Disadvantages of DYMO Protocol

The DYMO protocol presents a variety of new features over AODV. The performance evaluation shows that DYMO outperforms AODV as a MANET protocol. The advantages of the protocol can be summarized as follows:

- The protocol is energy efficient when the network is large and shows a high mobility.
- The routing table of DYMO is comparatively less memory consuming than AODV even with Path Accumulation feature.
- The overhead for the protocol decreases with increased network sizes and high mobility.

The DYMO protocol [7], however, does not perform well with low mobility. The control message overhead for such scenarios is rather high and unnecessary. Another limitation lies in the applicability of the protocol as stated in the DYMO Draft which states that DYMO performs well when traffic is directed from one part of the network to another. It shows a degraded performance when there is very low traffic random and routing overhead outruns the actual traffic.

## 4. SIMULATION ENVIRONMENT

All the simulations have been performed on Fedora 10 as the Operating System. Ns2.34 [1] has been installed on the platform for simulating the protocols along with add-on software such as Tracegraph etc. which is software for plotting graphs from the trace files. Ns2 implements ad hoc network protocols using traffic source behaviour such as FTP, CBR and VBR, routing algorithms such as Dijkstra, etc. It also supports multicasting in some of the routing protocols for LAN simulations [10]. The simulation tests have been performed on CBR traffic with varying number of nodes. Packet size has been set to 512 bytes and the pause time interval between transferring the packets has been set to 100 ms. Bit rate has been set to 2 Mbps with a Drop Tail of 15 ms. Simulations have been generated with the help of CMU traffic generator and the scenario with the help of setdest, which are the tools preinstalled with the ns2. The simulation field configuration has been set to 800 by 800 m.

The DYMO routing protocol has been simulated and analysed under varying pause time for changing network size against different performance metrics [2]. A network size of 40 nodes has been considered for simulations. Similarly data has been collected for other existing routing protocols such as AODV, DSDV and DSR just to represent the comparison between these in terms of various metrics and to study the performance of every protocol compared to DYMO. From the available set of performance metrics four have been considered in this study and simulations have been performed on all the protocols to calculate the effective values for metrics in each case.

1. **Packet delivery Fraction (PDF)**: It is the ratio of the amount of data packets delivered to the destination and total number of data packets sent by source.

2. **Average End-to-End Delay (AEED)**: The interval time between sending by the source node and receiving by the destination node, which includes the processing time and queuing time.

3. **Routing Overhead (RO)**: The total number of routing packets transmitted during simulation. Routing Overhead is important as it measures the scalability of a protocol, the degree to which it will function in congested or low bandwidth environments.

4. **Throughput (TP):** It is the average number of messages successfully delivered per unit time i.e. average number of bits delivered per second. Also refers to the amount of data transfer from source mode to destination in a specified amount of time.

## 5. SIMULATION RESULTS

Figures 3 to 6 show the graphical representation of the simulation values for all the four performance metrics with varying pause time and network size of 40 nodes.

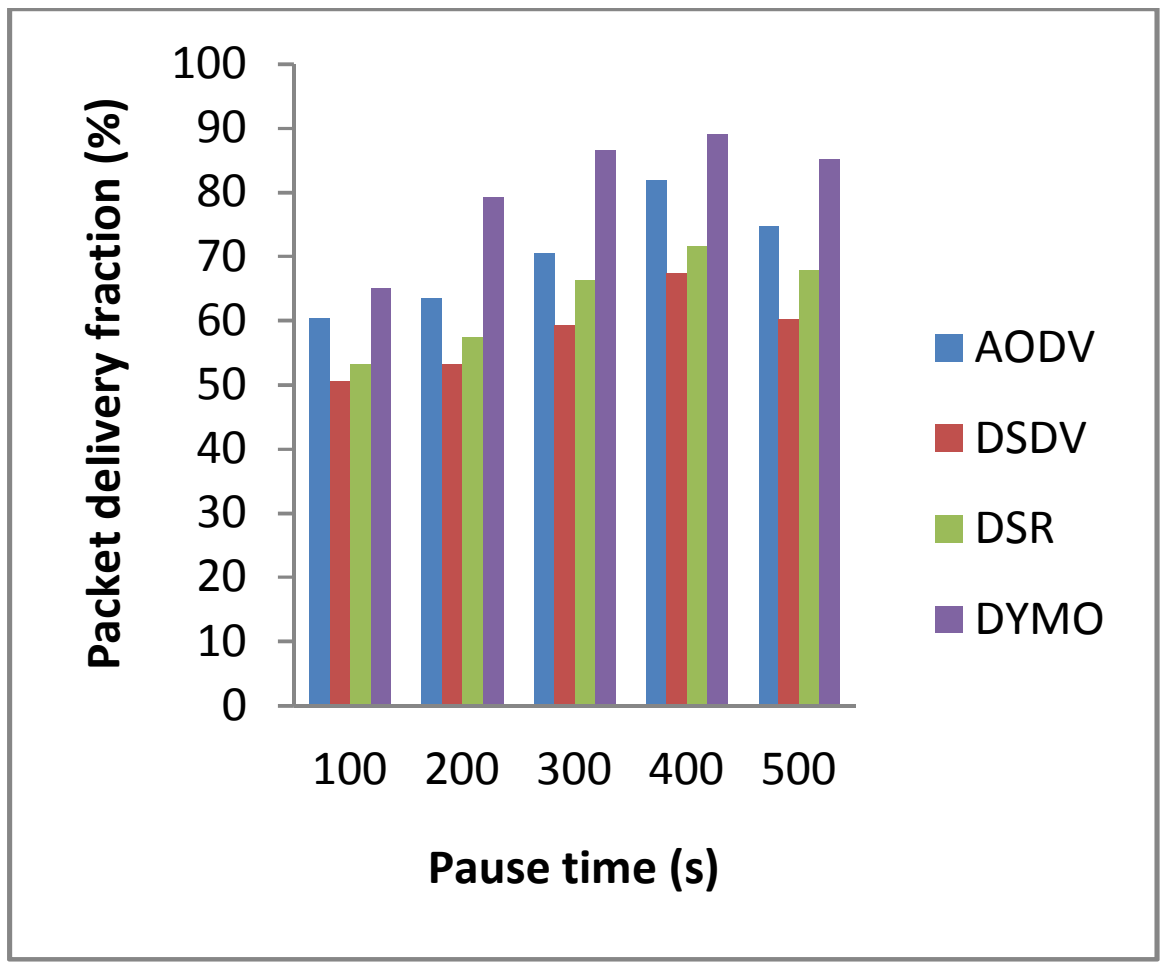

Figure 3. PDF with varying pause time

DYMO and AODV tend to have a higher packet delivery fraction which is a ratio of number of packets transmitted to number of packets dropped or lost whereas the packet delivery ratio of DSR and DSDV tend to be much lower than the other protocols. The performance of DSR protocol is better than DSDV because of the proper receiving of packets and less packet drop. But due to stale routes and more end to end delay it is observed that the performance of DSR protocol is not superior to DYMO and AODV. Due to the highest routing overhead in DSDV its packet delivery fraction is the worst as compared to the other protocols [8].

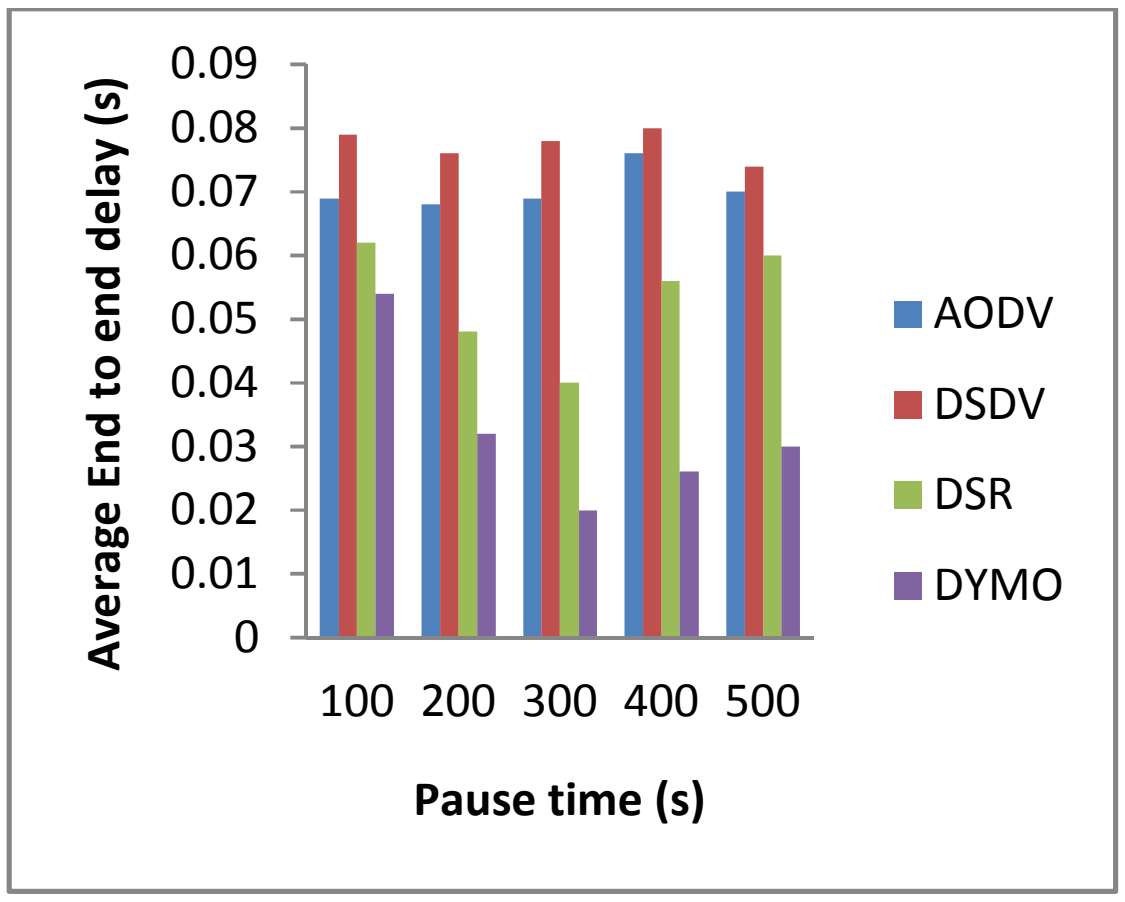

Figure 4. AEED with varying pause time

Average end to end delay of AODV Protocol is less than DSDV but more than DSR and DYMO for the reason that whenever a route from source to destination is required RREQ packets are sent to the neighbouring nodes of the source which further broadcast RREQ packets to their neighbouring nodes until a route to the required destination is not found. DSR exhibits lesser delay than AODV and DSDV because it caches every route it learns by receiving RREP. But due to the presence of stale caches in the whole scenario its overall delay is more than DYMO. Average end to end delay of DYMO protocol is the least due to the fact that each node needs not to start the route discovery process individually because path accumulation function helps each node to acquire knowledge about routes to other nodes.

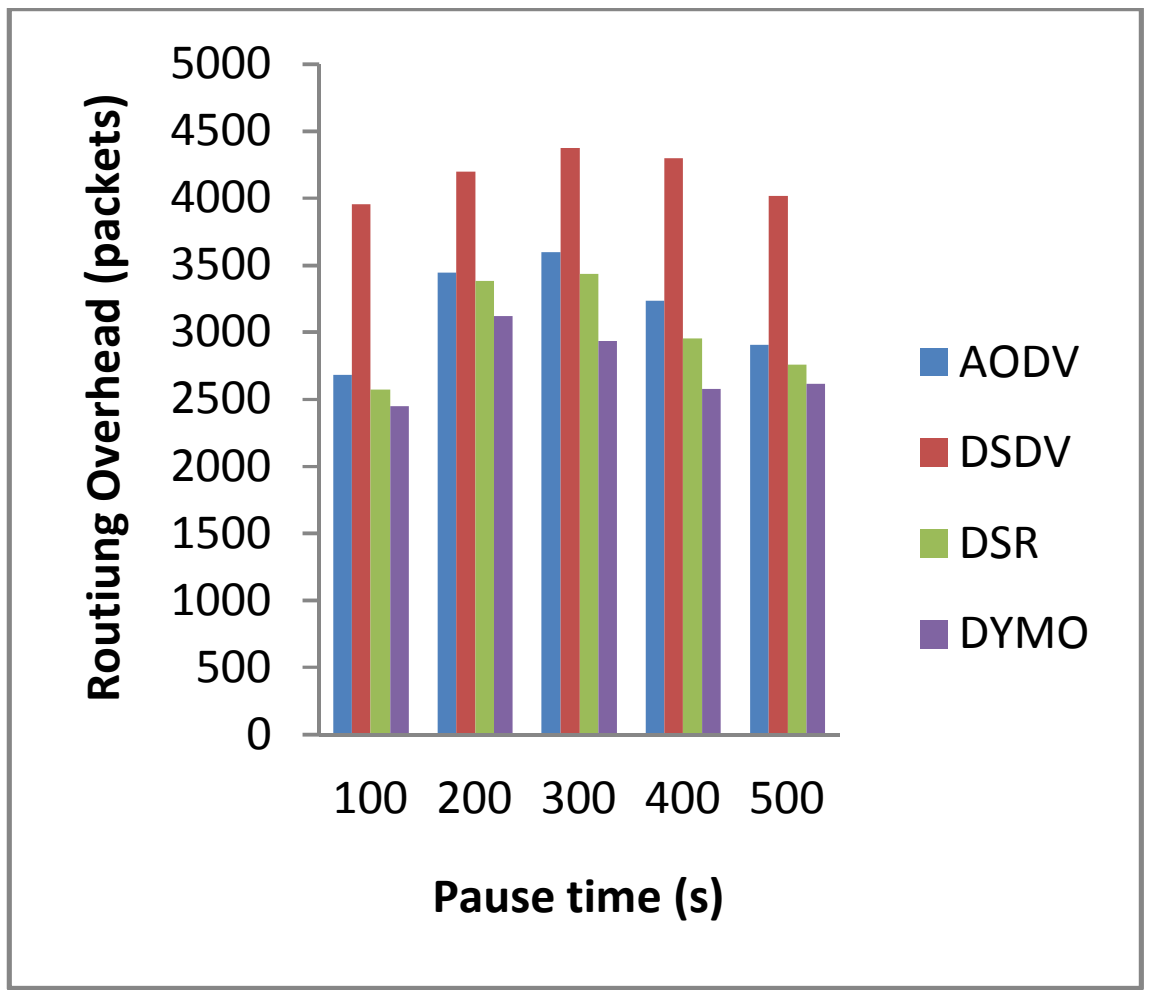

Figure 5. RO with varying pause time

Routing overhead in DYMO is the least because the route maintenance of DYMO is similar to that of AODV. But in DYMO the path accumulation function includes source routing characteristics, thereby allowing nodes listening to routing messages to acquire knowledge about routes to other nodes without initiating route request discoveries themselves. As a result, this path accumulation function leads to very less routing overhead as compared to other protocols.

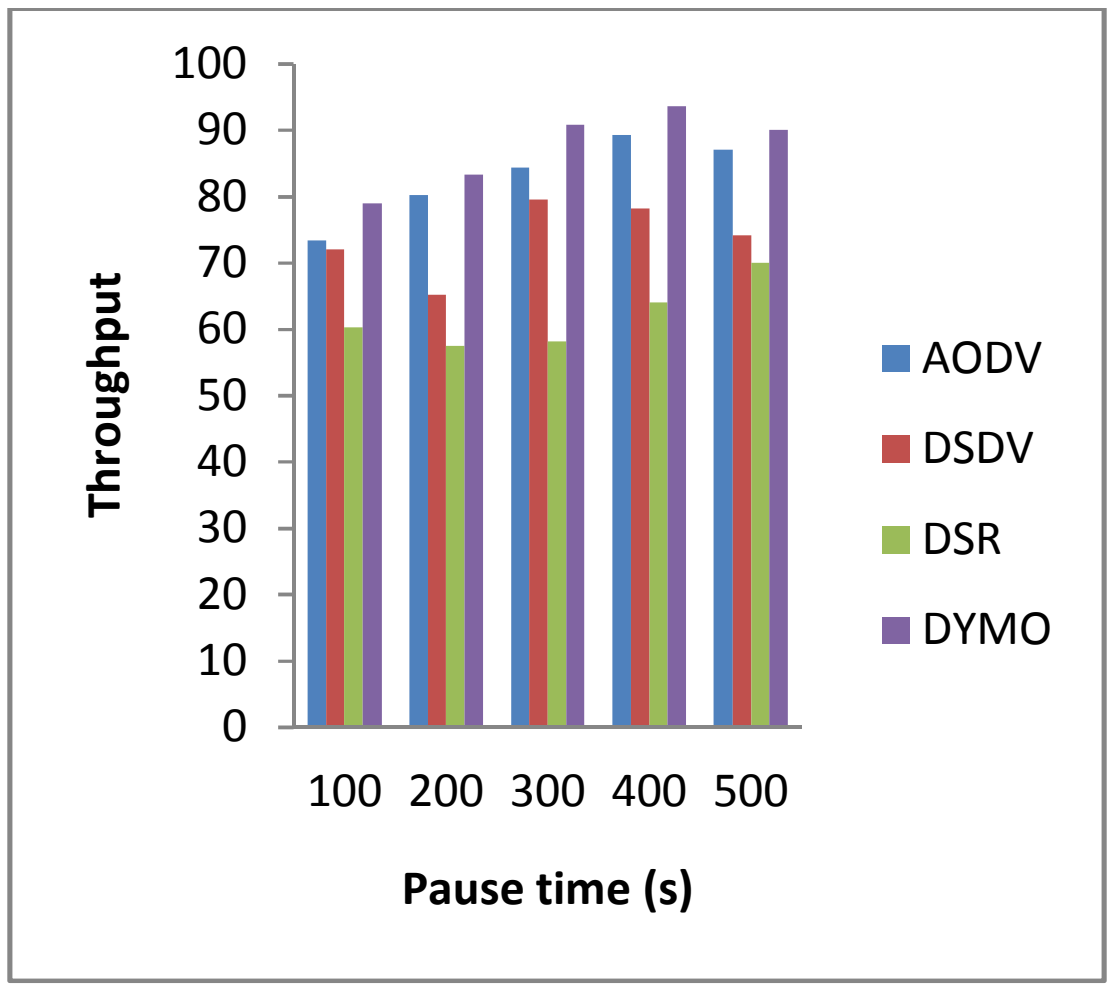

Figure 6. TP with varying pause time

DYMO exhibited highest throughput as compared to AODV, DSR and DSDV. Since more routing packets are generated and delivered by DYMO than other protocols. DYMO has ability to search route quickly as it avoids expiring good route by updating route lifetime appropriately. AODV shows higher throughput than the DSDV and DSR. AODV has much more routing packets than DSR because the AODV avoids loop and freshness of routes while DSR uses stale routes. DSDV protocol has better throughput than DSR Protocol, because of the proper receiving of packets and less packet drop in DSDV. Performance of DSR is weak as it doesn’t have proper technique to update stale routes and also in case of DSR simulation the packet loss is very high initially but it decreases substantially on the simulation time increases.

## 6. PERFORMANCE COMPARISON

Table 1 shows the overall comparison of all the four protocols simulated in section 3.7. It has been observed that DYMO has a high throughput and packet delivery, low average end to end delay but incurs a low routing overhead. On the other hand its processor AODV has shown a medium performance in case of all the metrics. DSR however generated a low values for all the metrics. And DSDV incurs a large routing overhead and delay thus having a low packet delivery and throughput values.

Table 1. Performance comparison of routing protocols

| Performance metrics | Routing Overhead | Throughput | Packet Delivery Fraction | Average End to End Delay |
|---|---|---|---|---|
| **High** | DSDV | DYMO | DYMO | DSDV |
| **Medium** | AODV | AODV | AODV | AODV |
| **Low** | DSR | DSDV | DSR | DSR |
| **Very low** | DYMO | DSR | DSDV | DYMO |

## 7. CONCLUSIONS

In this paper we have successfully simulated the existing DYMO routing protocol and analyzed its performance based on various simulation metrics. The simulation has been performed with varying pause times. It has been observed that DYMO being the successor of AODV performs better in all the terms. This paper will act as basis for many researchers worldwide to work upon the DYMO protocol and in future an effort will be done to enhance the performance of DYMO by using artificial intelligence techniques and simulations will be performed under varying network scenarios.

## AUTHORS

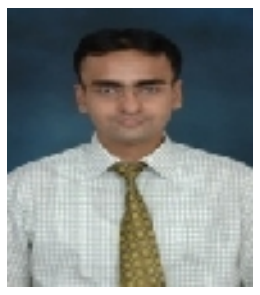

Anuj K. Gupta is a research fellow in Punjab Technical University, Punjab, India. His research area is Mobile ad hoc networks, wireless networks & Data Communication. He has a teaching experience of more than 10 years. Currently he is head of CSE Dept., RIMT Group.

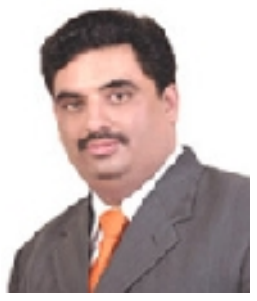

Dr. Harsh Sadawarti is Professor in Dept. of Computer Science & Engg. at RIMT, Punjab, India. He has a vast teaching and research experience of more than 20 years in the field of computer science. His areas of research are ad hoc networks, parallel computing & Distributed systems.

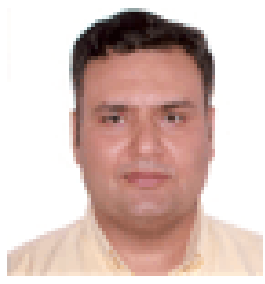

Dr. Anil K. Verma is faculty in Computer Sci. &Engg. Dept. at Thapar University, Patiala, Punjab, India. He has a vast teaching & research experience of more than 20 years. His areas of research are mobile ad hoc networks, wireless sensor networks & Network Security.